\documentstyle[prb,aps,twocolumn,epsfig,floats]{revtex}

\begin{document}

\draft

\twocolumn[\hsize\textwidth\columnwidth\hsize\csname 
@twocolumnfalse\endcsname

\date{\today}

\title{Evaluation of Exchange-Correlation Energy, Potential, and Stress}

\author{L.\ C.\ Balb\'as}
\address{
Departamento de F\'{\i}sica Te\'orica,
Universidad de Valladolid,
47011 Valladolid, Spain.}

\author{Jos\'e Lu\'{\i}s Martins}
\address{
Departamento de F\'{\i}sica, Instituto Superior T\'ecnico, 
Av Rovisco Pais, 1049-001 Lisboa, Portugal, and \\
INESC, Rua Alves Redol 9, Apartado 13069, 1000 Lisboa, Portugal.}

\author{Jos\'e M.\ Soler}
\address{
Departamento de F\'{\i}sica de la Materia Condensada,
Universidad Aut\'onoma de Madrid, 28049 Madrid, Spain.}

\maketitle

\begin{abstract}
   We describe a method for calculating the exchange and correlation (XC)
contributions to the total energy, effective potential, and stress tensor
in the generalized gradient approximation.
   We avoid using the analytical expressions for the functional derivatives
of $E_{\rm xc}(\rho)$, which depend on discontinuous second-order 
derivatives of the electron density $\rho$. 
   Instead, we first approximate $E_{\rm xc}$ by its integral in a real space 
grid, and then we evaluate its partial derivatives with respect to the
density at the grid points.
   This ensures the exact consistency between the calculated total energy,
potential, and stress, and it avoids the need of second-order derivatives.
   We show a few applications of the method, which requires only the value
of the (spin) electron density in a grid (possibly nonuniform) and returns 
a conventional (local) XC potential.
\end{abstract}


]

\narrowtext

\section{\bf Introduction}

   The generalized gradient approximation (GGA)\cite{GGA}
 to density functional
theory (DFT)\cite{HK} has been growing in acceptance in recent years,
due to the development of improved functionals and to the
realization of its higher accuracy, for many systems and 
properties, than that of the local density approximation (LDA)\cite{KS}.
   Some continuity problems of earlier functionals\cite{PW91,FUT} 
have been solved in more recent ones\cite{PBE}.
   Still, a basic problem is that, while $E_{\rm xc}^{\rm GGA}[\rho]$ depends 
only on the local value of $\rho({\bf r})$ and $|{\bf \nabla} \rho({\bf r})|$,
its functional derivative depends also on 
${\bf \nabla} | {\bf \nabla} \rho({\bf r})|$, which is discontinuous where
${\bf \nabla} \rho({\bf r}) = 0$.
   This implies that, to avoid aliasing effects, very fine integration 
grids are required to evaluate the XC potential $v_{\rm xc}({\bf r})$
and its matrix elements.
   A different problem occurs with grid-oriented implementations
of DFT\cite{Chelikowsky,Hamann,Kaxiras,Bernholc,Beck}, in which the electron
density is known only at the grid points, while its gradients must
be evaluated using finite differences.
   In this case, an inconsistency between the energy $E_{\rm xc}$ and 
the potential $v_{\rm xc}$ may occur because different formulas need to 
be used for the higher-order derivatives of the density.
   Hamann\cite{Hamann} proposed an elegant solution to these problems
by defining a nonlocal potential, which operates on the {\em gradient}
of the electron wave functions.
   Although in principle this does not pose any special difficulty, 
it requires a specific, unconventional implementation, which may be 
difficult to adapt to existing DFT codes.
   White and Bird\cite{WB} found another solution, by {\em defining}
$E_{\rm xc}^{\rm GGA}$ as an integral in a real space grid (what is always
done in practice), and $v_{\rm xc}({\bf r}_i)$ as its {\em partial}
derivative (as opposed to functional derivative) with respect to
the density $\rho_i$ at the grid points ${\bf r}_i$.
   They applied this method to a plane wave basis set and a uniform 
integration grid, finding the gradients with fast Fourier transforms 
(FFTs).
   Here we generalize their method to arbitrary bases and nonuniform 
grids, and we calculate the density gradient using finite differences.
   The method produces a standard, local potential, which is exactly 
consistent with the definition of $E_{\rm xc}$, in the sense that $v_{\rm xc}$
is the correct potential in the Schr\"odinger equation that results 
from the variational minimization of the total energy.
   In addition, we show how to evaluate the XC contribution to the 
stress tensor in crystals, following the same ideas.
   The trivial particular case of the LDA is also discussed.

\section{\bf Exchange-Correlation Energy and Potential}

   The LDA\cite{KS} and GGA\cite{GGA} approximations to $E_{\rm xc}$ have 
the general forms
$$
   E_{\rm xc}^{LDA}[\rho] = \int \! f_{LDA}
       (\rho({\bf r})) \, d{\bf r},
$$
\begin{equation}
   E_{\rm xc}^{\rm GGA}[\rho] =
       \int \! f_{\rm GGA}
       (\rho({\bf r}),{\bf g}({\bf r})) \, d{\bf r},
   \label{eqggadef}
\end{equation}
where $f = \rho~\epsilon_{\rm xc}$ is the local XC 
energy density, and $\epsilon_{\rm xc}$ is the XC energy per electron. 
   We use the notation \cite{notation}
${\bf g}({\bf r}) \equiv {\bf \nabla} \rho({\bf r})$,
$g_\alpha({\bf r}) \equiv \nabla_\alpha \rho({\bf r})$, and
$g({\bf r}) \equiv |{\bf \nabla} \rho({\bf r})|$.
   The argument ${\bf g}({\bf r})$ in Eq.(\ref{eqggadef}), 
rather than $g({\bf r})$, indicates that, in principle, 
$f_{\rm GGA}$ might depend on the relative 
orientation of the gradients of the two spin components of the density.
   However, for notational simplicity we will omit the spin index and
the sum over it, since their inclusion is trivial.
   In the limit of slowly varying electron density, we must recover 
the LDA result, that is 
$f_{\rm GGA}(\rho,0) = f_{\rm LDA}(\rho)$\cite{FUT}.

   In the LDA the XC potential has a simple expression,
$$
   v_{\rm xc}^{\rm LDA} = {d f_{\rm LDA} \over d\rho} =
        \epsilon_{\rm xc}^{\rm LDA} +
     \rho {d \epsilon_{\rm xc}^{\rm LDA} \over d \rho},
$$
whereas in the GGA it is considerably more complicated
\begin{eqnarray}
   v_{\rm xc}^{\rm GGA} & = &
        {\partial f_{\rm GGA} \over \partial\rho}
          - {\bf \nabla}  {\partial f_{\rm GGA}
                              \over \partial {\bf g}} \nonumber\\
   & = & \epsilon_{\rm xc}^{\rm GGA} +
     \rho {\partial \epsilon_{\rm xc}^{\rm GGA} \over \partial \rho}
     - g{\partial \epsilon_{\rm xc}^{\rm GGA} \over
                                   \partial g}
     - \rho g {\partial^2 \epsilon_{\rm xc}^{\rm GGA} \over
                        \partial g\partial \rho} \nonumber \\
   & & - {\rho \over g} {\partial \epsilon_{\rm xc}^{\rm GGA} \over
                         \partial g} \nabla^2 \rho
       + {\rho \over g^2}{\partial \epsilon_{\rm xc}^{\rm GGA} \over
            \partial g} {\bf g} {\bf \nabla} g
       \nonumber \\
   & & - {\rho \over g}{\partial^2 \epsilon_{\rm xc}^{\rm GGA} \over
                \partial g^2} {\bf g} {\bf \nabla} g,
   \label{eqvxcgga}
\end{eqnarray}
as it depends on the first and second gradients of the electron density.
   Since $|{\bf \nabla} \rho|$ has cusps at extrema of $\rho$, ${\bf \nabla} g$
is discontinuous at these points, what causes problems for its numerical 
representation.
   Furthermore, some parameterizations of $\epsilon_{\rm xc}^{\rm GGA}$
which apparently join seamlessly different density regimes,
may have higher derivatives which do not behave well
in those transition regions.

   In what follows, we will omit the label GGA, except to underline a 
distinction with LDA.
   In a practical calculation, the XC energy is calculated through a 
numerical integration. 
   From a set of $M$ mesh points ${\bf r}_i$ and weights $w_i$, we approximate
$$ 
   E_{\rm xc} \simeq 
   \sum_{i=1}^M w_i f(\rho({\bf r}_i),{\bf g}({\bf r}_i)).
$$
   The weights may be, for example $4 \pi r^2_i \Delta r$ in a uniform radial 
grid, or the jacobian of the local metric tensor in an adaptive-coordinate 
grid\cite{Gygi} (see below).
   In addition, we must specify precisely the meaning of 
${\bf g}({\bf r}_i)$.
   If a well defined basis set is used, the electron density and its gradient
can be calculated exactly at any point in space, from the electron 
wave functions and their gradients.
   In practice, this may add an appreciable overhead in terms of 
computer time and memory.
   Alternatively, we can use the values of the density at the grid points
to calculate its gradient, using either FFTs\cite{WB} or finite differences.
   As the gradient is a linear operator, we can write in general
\begin{equation}
   g_\alpha({\bf r}_i) \simeq 
   g_{i \alpha} \equiv \sum_{j=1}^M A_{ij}^\alpha \rho_j,
   \label{gi}
\end{equation}
and $\partial g_{i \alpha} / \partial \rho_j = A_{ij}^\alpha$.
$\{i,j\}$ are combined indexes ($i \equiv \{i_1,i_2,i_3\}$)
that label grid points, $\alpha$ labels the three cartesian
coordinates, $\rho_i \equiv \rho({\bf r}_i)$,
and the coefficients $A_{ij}^\alpha$ depend on the mesh 
and the chosen numerical derivative formula, but not on $\rho_i$.
   There are many choices for the coefficients $A_{ij}^\alpha$,
depending on the integration mesh and interpolation method,
since the numerical derivative can be defined as the derivative of the
interpolation function.
   The particular case described by White and Bird\cite{WB} uses a uniform
mesh and a Fourier-series interpolation. 
   It was developed for plane-wave calculations,
and FFTs were used to evaluate Eq.~(\ref{gi}).
   Our method can be generalized to nonuniform grids.
   In the implementation that we will
describe later, we use a local Lagrange interpolation, for which
the matrix $A_{ij}^\alpha$ is sparse.

   Notice that in Eq.~(\ref{gi}) we are {\em defining} ${\bf g}_i$ as
the numerical derivative of $\rho$ on the mesh
(therefore being a function of the values $\{\rho_j\}$),
while we reserve the notation ${\bf g}({\bf r}_i)$ for the exact
gradient of $\rho$ at ${\bf r}_i$ (in case it is known).
   It may be argued that the use of ${\bf g}_i$, instead of 
${\bf g}({\bf r}_i)$ represents an additional approximation.
   However, in practice ${\bf g}_i$ is frequently a better approximation 
than ${\bf g}({\bf r}_i)$
to the {\em average} value of ${\bf g}({\bf r})$ within the spatial 
``pixel'' which corresponds to the integration point ${\bf r}_i$.
   Both values agree for a Fourier interpolation, provided that the plane wave 
cutoff of the grid is twice as large as that of the wave functions.
   And of course they must also agree, for any interpolation scheme, in the
limit of an infinitely fine grid.

   Thus, following White and Bird\cite{WB}, we {\em define} 
\begin{equation}
   {\tilde E}_{\rm xc}(\{\rho_i\}) \equiv
   \sum_{i=1}^M w_i f_i(\rho_i,{\bf g}_i(\{\rho_j\})) \simeq 
   E_{\rm xc},
   \label{eqExcgrid}
\end{equation}
and
\begin{eqnarray} 
   {\tilde v}_i^{\rm xc}
    & \equiv & {1 \over w_i}
      {\partial  {\tilde E}_{\rm xc}(\{\rho_j\}) \over \partial \rho_i}
      \nonumber \\
    & = & {\partial f \over \partial \rho_i} +
         \sum_{j=1}^M \sum_{\alpha = 1}^3 {w_j \over w_i}
           {\partial f_j \over \partial g_{j \alpha}}
                    {\partial  g_{j \alpha} \over \partial \rho_i}.
   \label{eqWB}
\end{eqnarray}
   It is important to emphasize that Eqs.~(\ref{eqvxcgga}) and (\ref{eqWB})
are asymptotically equivalent in the limit of an infinitely fine
grid, but different in practice
(in particular, Eq.~(\ref{eqWB}) has no explicit dependence on the 
second derivatives of $\rho$).
   In fact, it is easy to see that Eq.~(\ref{eqWB}), and not 
Eq.~(\ref{eqvxcgga}), is the ``correct'' definition of $v_{\rm xc}$, 
if the functional form (\ref{eqExcgrid}) is actually used in
the variational energy minimization, 
$\partial E / \partial \psi^* = H \psi,$
because, from Eq.~(\ref{eqWB}),
$$ 
   {\partial {\tilde E}_{\rm xc} \over \partial \psi_i^*} =
   {\partial {\tilde E}_{\rm xc} \over \partial \rho_i} 
   {\partial \rho_i \over \partial \psi_i^*} =
   w_i {\tilde v}_i^{\rm xc} \psi_i.
$$

   In the LDA, $f_i$ does not depend on ${\bf g}_i$, and only the first
term in Eq.~(\ref{eqWB}) remains, giving the trivial result
$$ 
   {\tilde v}_i^{\rm LDA} = {df_i^{\rm LDA} \over d\rho_i}.
$$

   There remains the problem of determining the coefficients $A_{ij}^\alpha$
and the weights $w_i$ in Eqs.~(\ref{gi}) and (\ref{eqExcgrid}).
   To this purpose, it is convenient to introduce grid variables 
$\{s_\mu; \mu=1,2,3\}$, which are in principle continuous.
   In practice, however, the density $\rho({\bf s})$ and the cartesian 
coordinates ${\bf r}({\bf s})$ of the grid points are evaluated only at 
integer values of $s_\mu$.
   For a regular grid, 
$r_\alpha({\bf s}) = \sum_{\mu=1}^3 s_\mu a_{\mu \alpha} / N_\mu$,
where $a_{\mu \alpha}$ is the $\alpha$th cartesian coordinate of the
$\mu$th lattice unit vector, and $N_\mu$ is the number of
grid divisions along that vector.
   For nonuniform grids, $r_\alpha({\bf s})$ are general
functions and it is convenient to introduce the jacobian of the 
transformation ${\bf s} \rightarrow {\bf r}$, 
   The XC energy can then be expressed as
$$
   E_{\rm xc} = \int f(\rho({\bf s}), {\bf g}({\bf s})) 
       \left| \frac{\partial {\bf r}}{\partial {\bf s}} \right| d{\bf s} 
$$
or
$$
   {\tilde E}_{\rm xc} = \sum_{i=1}^M f(\rho_i, {\bf g}_i) 
                  \left| D_i \right|
$$
where we have used that $\Delta s_\mu = 1$ by definition of $s_\mu$.
$\left| D_i \right|$ is the determinant at point $i$ of
the matrix of partial derivatives
$$
   D_{\alpha \mu}({\bf s}) \equiv 
   \frac{\partial r_\alpha({\bf s})}{\partial s_\mu}.
$$
   By comparison with (\ref{eqExcgrid}), we conclude that 
$w_i = \left| D_i \right|$.
   Since the grid variables $s_\mu$ are by definition ``orthogonal''
and the functions $r_\alpha$ and $\rho$ are evaluated at regular unit 
intervals of $s_\mu$, their derivatives $\partial/\partial s_\mu$ can be
calculated straightforwardly: 
\begin{equation} 
D_i^{\mu \alpha} =
   \left( \frac{\partial r_\alpha}{\partial s_\mu} \right)_i =
   \sum_{j=1}^M B_{ij}^\mu r_{j \alpha}
\label{Di}
\end{equation}
$$ 
   \left( \frac{\partial \rho}{\partial s_\mu} \right)_i =
   \sum_{j=1}^M B_{ij}^\mu \rho_j 
$$
where the coefficients $B_{ij}^\mu$ depend only on the relative
values of $i$ and $j$, and are independent of the grid 
coordinates ${\bf r}({\bf s})$:
$$
   B^\mu_{ij} = \left\{ 
   \begin{array}{ll}
      L_k^{(n)} & \mbox{if $j_\mu=i_\mu+k; j_\nu=i_\nu, \nu \neq \mu$}  \\
     -L_k^{(n)} & \mbox{if $j_\mu=i_\mu-k; j_\nu=i_\nu, \nu \neq \mu$}  \\
       0      & \mbox{otherwise} 
   \end{array} \right.
$$
where the coefficients $L_k^{(n)}$ may be derived from a
$(2n+1)$-point Lagrange polynomial interpolation formula~\cite{AS}.

   We can now calculate ${\bf g}$ as
$$
    g_{i \alpha} = \sum_{\mu=1}^3 
        \left( \frac{\partial \rho}{\partial s_\mu}  \right)_i
        \left( \frac{\partial s_\mu}{\partial r_\alpha} \right)_i
      = \sum_{\mu=1}^3 \sum_{j=1}^M 
        (D^{-1}_i)_{\alpha \mu} B_{ij}^\mu \rho_j 
$$
and by identifying coefficients with Eq.~(\ref{gi})
\begin{equation}
   A_{ji}^\alpha = {\partial g_{j \alpha} \over \partial \rho_i} = 
   \sum_{\mu=1}^3 (D^{-1}_j)_{\alpha \mu} B_{ji}^\mu. 
   \label{dgdrho}
\end{equation}
   Thus, the calculation of the XC potential from the density on the 
grid involves the following steps, for every point $j$:
1) find the $3 \times 3$ matrix $D_j$, using Eq.~(\ref{Di}),
   its determinant $w_j$, and its inverse $D_j^{-1}$,
   storing $w_j$ for later use;
2) find $A_{ji}^\alpha$ from Eq.~(\ref{dgdrho})
   and ${\bf g}_j$ from Eq.~(\ref{gi});
3) calculate $f(\rho_j,{\bf g}_j)$ and its derivatives with
   respect to $\rho_j$ and $g_{j \alpha}$;
4) add the first term in Eq.~(\ref{eqWB}) (multiplied by $w_j$) to 
   ${\tilde v}_j^{\rm xc}$ and the second term (except the denominator $w_i$)
   to ${\tilde v}_i^{\rm xc}$ for every neighbor point $i$ involved in the 
   calculation of ${\bf g}_j$;
5) when the previous loop is finished, run again over all grid points 
   $i$, dividing ${\tilde v}_i^{\rm xc}$ by $w_i$.
   In the case of a uniform grid, the matrix $D_i$ does not depend on $i$. 
   Thus, $w_i$ and $D_i^{-1}$ can be evaluated once and for all, saving 
steps 1 and 5, as well as the temporary array required to store $w_i$.

\section{\bf Exchange-Correlation Contribution to the Stress Tensor}

   We consider now the stress tensor \cite{sign}
$$
   \sigma^{\rm xc}_{\alpha \beta} \equiv 
     {\partial {\tilde E}_{\rm xc} \over \partial \epsilon_{\alpha \beta}}
$$
where $\epsilon_{\alpha \beta}$ is the strain tensor, giving the deformation
of all points in space (including atomic and grid coordinates):
\begin{equation}
   r_\alpha \rightarrow \sum_{\beta=1}^3 (\delta_{\alpha \beta} + 
   \epsilon_{\alpha \beta}) r_\beta.
   \label{rmu}
\end{equation}
   More generally, we consider the derivative of ${\tilde E}_{\rm xc}$ with 
respect to a parameter $\lambda$ that affects the position of the 
grid points.
   This may be $\epsilon_{\alpha \beta}$ or one of the atomic positions, 
in case of a nonuniform grid which depends on them \cite{Gygi}.
   It is therefore convenient to recognize explicitly that, 
when the system is modified, ${\tilde E}_{\rm xc}$ depends on the 
grid point coordinates in addition to the densities at the grid points, i.e.\
${\tilde E}_{\rm xc}( \{ \rho_i \}, \{ {\bf r}_i \} )$.
   Thus, 
$$
   {\partial {\tilde E}_{\rm xc} \over \partial \lambda} =
   \sum_{i=1}^M \left( {\partial {\tilde E}_{\rm xc} \over \partial \rho_i} 
         {\partial \rho_i \over \partial \lambda} +
   \sum_{\alpha=1}^3 {\partial {\tilde E}_{\rm xc} \over \partial r_{i \alpha}}
               {\partial r_{i \alpha} \over \partial \lambda} \right)
$$
where $\partial {\tilde E}_{\rm xc} / \partial \rho_i = 
{\tilde v}_i^{\rm xc}$ and
$$
{\partial {\tilde E}_{\rm xc} \over \partial r_{i \alpha}} =
   \sum_{j=1}^M \left( 
   {\partial |D_j| \over \partial r_{i \alpha}} f_j +
   |D_j| \sum_{\beta=1}^3 {\partial f_j \over \partial g_{j \beta}}
          {\partial g_{j \beta}   \over \partial r_{i \alpha}} \right)
$$
$$
   {\partial g_{j \beta} \over \partial r_{i \alpha}} = \sum_{\mu=1}^3 
   \left( {\partial \rho \over \partial s_\mu} \right)_j
              {\partial \over \partial r_{i \alpha}} (D^{-1}_j)_{\beta \mu}
$$
   Now, using Eq.~(\ref{rmu}) we find 
\begin{equation}
{\partial r_\gamma \over \partial \epsilon_{\alpha \beta}} =
   \delta_{\gamma \alpha} r_\beta,
\label{drdeps}
\end{equation}
and taking into account the general properties of a determinant,
as well as the fact that 
$(D D^{-1} = I) \Rightarrow (\delta D = - D^{-1} \delta D D^{-1})$, 
we finally find
\begin{equation}
   {\partial {\tilde E}_{\rm xc} \over \partial \epsilon_{\alpha \beta}} = 
   \delta_{\alpha \beta} {\tilde E}_{\rm xc} +
   \sum_{i=1}^M w_i {\tilde v}_i^{\rm xc} 
    {\partial \rho_i \over \partial \epsilon_{\alpha \beta}} -
   \sum_{i=1}^M w_i 
    {\partial f_i \over \partial g_{i \beta}} g_{j \alpha}
   \label{dEdeps}
\end{equation}
   In the limit of an infinitely fine grid, we have
\begin{eqnarray}
   {\partial E_{\rm xc} \over \partial \epsilon_{\alpha \beta}} & = &
   \delta_{\alpha \beta} E_{\rm xc} +
   \int v_{\rm xc}({\bf r}) 
   {\partial \rho({\bf r}) \over \partial \epsilon_{\alpha \beta}} 
   d{\bf r} \nonumber \\
   & & - \int {\partial f(\rho({\bf r}),{\bf g}({\bf r})) \over 
        \partial g_\beta({\bf r})} g_\alpha({\bf r}) d{\bf r}.
   \label{dEdepsInt}
\end{eqnarray}

   The presence of $\partial \rho / \epsilon_{\alpha \beta}$ in 
Eqs.~(\ref{dEdeps}) and (\ref{dEdepsInt}) requires to clarify what is 
kept constant when taking the derivatives.
   We will argue that the ``correct'' definition (or at least the most 
useful one) requires to keep constant the variational coefficients $c_{na}$
of the expansion of the electron wave functions $\psi_n({\bf r})$ in
terms of the $N$ basis functions $\phi_a({\bf r})$:
\begin{equation}
   \psi_n({\bf r}) = \sum_{a=1}^N c_{na} \phi_a({\bf r}) 
   \label{psin}
\end{equation}
   Since $\sigma_{\alpha \beta}$ is generally evaluated at the electronic 
ground state, the Hellman-Feynman theorem ensures that the change of $c_{na}$
will not affect the total energy to first order.
   Of course, the Hellman-Feynman theorem does not apply to $E_{\rm xc}$
alone, but in practice we are interested in adding 
$\sigma^{\rm xc}_{\alpha \beta}$ to other contributions of 
$\sigma_{\alpha \beta}$, in order to calculate the total stress tensor.
   And even in the Car-Parrinello~\cite{CarParrinello} method, in which
forces and stresses are evaluated out of the ground state, the appropriate
definition of these magnitudes involves the derivatives of the total energy
at constant $c_{na}$.

   There are two reasons why $\rho_i$ depends on $\epsilon_{\alpha \beta}$:
the change of the basis functions $\phi_a({\bf r}_i)$ at the displaced 
grid points, and the change of the wave function coefficients $c_{na}$
required to maintain the orthonormality constraints 
$\langle \psi_n | \psi_m \rangle = \delta_{nm}$.
   With a plane wave basis set, or with a grid-oriented 
scheme~\cite{Chelikowsky,Bernholc,Kaxiras}, in which the wave functions
are defined directly at the grid points, without any basis, the 
orthonormality constraints are not affected by the deformation of the
unit cell, provided that the wave functions and the density are simply 
scaled by a factor $(\Omega_0/\Omega)^{1/2}$ and $\Omega_0/\Omega$, 
respectively, where $\Omega_0$ and $\Omega$ are the unit cell volumes 
before and after the deformation.
   It is easy to verify that 
$\partial \Omega / \partial \epsilon_{\alpha \beta} = 
\Omega \delta_{\alpha \beta}$
and therefore
\begin{equation}
   {\partial \rho({\bf r}) \over \partial \epsilon_{\alpha \beta}} = 
   - \delta_{\alpha \beta} \rho({\bf r})
   \label{drhodepsPW}
\end{equation}
   Substitution of (\ref{drhodepsPW}) into (\ref{dEdepsInt}) leads to
\begin{eqnarray}
   {\partial E_{\rm xc} \over \partial \epsilon_{\alpha \beta}} 
   & = &
       \delta_{\alpha \beta} \int 
       \left( \epsilon_{\rm xc}({\bf r}) - 
               v_{\rm xc}({\bf r}) \right) \rho({\bf r}  
          d{\bf r} \nonumber \\
   &  & - \int {\partial f(\rho({\bf r}),{\bf g}({\bf r})) \over 
         \partial g_\beta({\bf r})} g_\alpha({\bf r}) d{\bf r}, \nonumber
\end{eqnarray}
which coincides with Eq.~(24) of Dal Corso and Resta\cite{CR}.

   In the case of an atomic basis set, it is convenient to define the
density matrix
$$
   \rho_{ab} = \sum_{n=1}^N q_n c^*_{na} c_{nb}
$$
with $q_n$ the occupation of state $\psi_n$.
Then
$$
   \rho({\bf r}) = \sum_{a,b=1}^N \rho_{ab} \phi_a({\bf r}) \phi_b({\bf r})
$$
\begin{eqnarray}
   {\partial \rho({\bf r}) \over \partial \lambda} & = & 
      \sum_{a,b=1}^N \left( {\partial \rho_{ab} \over \partial \lambda}
             \phi_a({\bf r}) \phi_b({\bf r}) \right. \nonumber \\
   & & \left. + 2 \rho_{ab} \phi_a({\bf r}) \sum_{\alpha=1}^3 
   \nabla_\alpha \phi_b({\bf r})
   {\partial (r_\alpha - R_{b \alpha}) \over \partial \lambda} \right)
   \label{drhodeps}
\end{eqnarray}
where we are assuming real basis orbitals for simplicity.
${\bf R}_b$ is the origin (atomic position) of orbital $\phi_b$, and
the last term in Eq.~(\ref{drhodeps}) accounts for the change in the relative
position ${\bf r}-{\bf R}_b$ when we deform the lattice or change the
grid point positions (we assume a constant shape of $\phi_b$). 
   $\partial \rho_{ab} / \partial \lambda$ is the change in the
density matrix required to maintain the orthonormality constraints:
\begin{equation}
   {\partial \rho_{ab} \over \partial \lambda} = 
   - \sum_{c=1}^N \rho_{ac} {\partial S_{cb} \over \partial \lambda} =
   - \sum_{c=1}^N \rho_{ac} \sum_{\alpha=1}^3 
     {\partial S_{cb} \over \partial r_{cb}^\alpha}
     {\partial r_{cb}^\alpha \over \partial \lambda}
   \label{drhodlambda}
\end{equation}
where $S_{ab} \equiv \langle \phi_a | \phi_b \rangle$, and 
${\bf r}_{ab} \equiv {\bf R}_b - {\bf R}_a$.
   By combining Eqs.~(\ref{drdeps}), (\ref{drhodeps}), and 
(\ref{drhodlambda}) we obtain
\begin{eqnarray}
   \int v_{\rm xc}({\bf r}) 
      {\partial \rho({\bf r}) \over \partial \epsilon_{\alpha \beta}}
      & & d{\bf r} =  - \sum_{a,b,c=1}^N
       \rho_{ac} {\partial S_{cb} \over \partial r_{cb}^\alpha} r_{cb}^\beta
        \langle \phi_a | v_{\rm xc} | \phi_b \rangle  \nonumber \\
      & & + 2 \sum_{a,b=1}^N \rho_{ab}
      \langle \
         \phi_a | v_{\rm xc}({\bf r}) (r_\beta-R_{b \beta}) \nabla_\alpha | \phi_b 
      \rangle. \nonumber
\end{eqnarray}
   The same expression is valid if the integral is replaced by the grid sum 
$\sum_i w_i {\tilde v}_i^{\rm xc} 
(\partial \rho_i/\partial \epsilon_{\alpha \beta})$,
except that then the matrix elements must also be calculated on the grid, i.e.
$$
   \langle \phi_a | v_{\rm xc} | \phi_b \rangle = 
      \sum_{i=1}^M w_i {\tilde v}^{\rm xc}_i \phi_{a i} \phi_{b i}
$$
   Similar terms appear in other contributions to the stress tensor.
   Thus, in the derivation of the Hartree energy, there is an
identical term, except for the substitution of $v_{\rm xc}$ by the Hartree
potential $v_H$.
   Then the total contribution to the stress tensor that arises from
the change in the density has the form
\begin{eqnarray}
   \int {\delta E \over \delta \rho({\bf r})}
     & & {\partial \rho({\bf r}) \over \partial \epsilon_{\alpha \beta}} 
       d{\bf r} =   
     - \sum_{a,b=1}^N E_{ab} {\partial S_{ab} \over \partial r_{ab}^\alpha} 
       r_{ab}^\beta \nonumber \\
     & & + 2 \sum_{a,b=1}^N \rho_{ab} 
      \langle 
        \phi_a | v({\bf r}) (r_\beta-R_{b \beta}) \nabla_\alpha | \phi_b 
      \rangle
   \label{dvrhodeps}
\end{eqnarray}
where $v({\bf r})$ is the total effective potential, and $E_{ab}$ are 
energy-density matrix elements,
\begin{equation}
   E_{ab} = \sum_{c=1}^N \rho_{ac} H_{cb} =
      \sum_{n=1}^N q_n \epsilon_n c^*_{na} c_{nb},
   \label{Eab}
\end{equation}
with $H_{ab} = \langle \phi_a | H | \phi_b \rangle$, $H$ the total 
one-electron hamiltonian and $\epsilon_n$ its $n$th eigenvalue.
   Notice, however, that the derivation of Eqs.~(\ref{dvrhodeps}) and 
(\ref{Eab}) does {\em not} assume that the matrix elements $S_{ab}$ 
and $H_{ab}$ or their derivatives are calculated on the grid.
   In fact, in our atomic-basis DFT implementation~\cite{siesta}, 
two-center matrix elements like $S_{ab}$ or 
$\langle \phi_a | -\nabla^2 | \phi_b \rangle$, are calculated using
reciprocal-space convolution techniques~\cite{Sankey}.

\section{\bf Accuracy tests}

   We have implemented the method described above as an independent package,
which we are using with several electronic structure programs,
including an atomic pseudopotential generator, the localized-basis
program {\sc Siesta}~\cite{siesta} and a pseudopotential plane-wave program.

   We have implemented two different LDA parameterizations,
PZ-LDA\cite{PZ} and PW92-LDA\cite{PW91},
of Ceperley and Adler electron gas energies\cite{CA} for $f_{\rm LDA}$
and two different GGA recipes, PW91-GGA\cite{PW91} and
PBE-GGA\cite{PBE} for $f_{\rm GGA}$.
   They were all implemented in their spin-dependent formulations.
   As the calculation of $f$ and its derivatives are kept as
separate procedures it is easy to add new parameterizations.
   As only first derivatives of $f$ are needed in our method
to calculate stresses and potentials, we do not have to calculate
second derivatives as in the traditional method. 
   Since GGA functional forms tend to be complicated, this is convenient.

   In the atomic pseudopotential generation program the electron density
is assumed to have spherical symmetry.
   The radial mesh points are derived from a monotonous function $r(s)$. 
   The gradient of the density is calculated analogously to the 
three-dimensional case:
$$
   g_i = \left( {dr \over ds} \right)_i^{-1}
         \left( {d\rho \over ds} \right)_i,
$$
$$
   \left( {dr \over ds} \right)_i = 
      \sum_{k=1}^n L_k^{(n)} (r_{i+k} - r_{i-k}),
$$
$$
   \left( {d\rho \over ds} \right)_i = 
      \sum_{k=1}^n L_k^{(n)} (\rho_{i+k} - \rho_{i-k}).
$$

\begin{figure}
\epsfig{figure=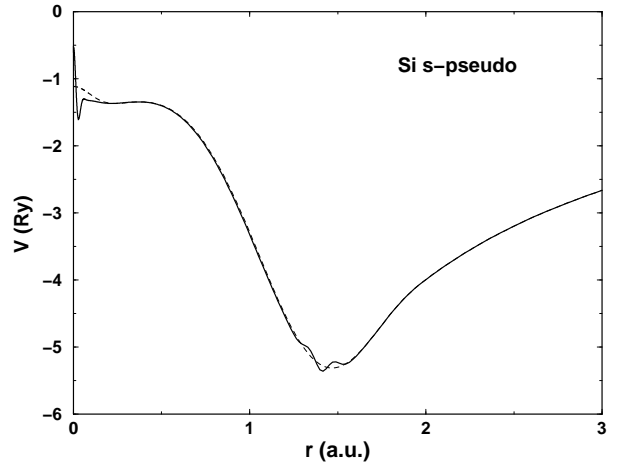,width=85mm}
\caption{
   $s$-pseudopotential generated for Si.
   The solid curve was generated with the PW91-GGA and the dashed curve
with the PBE-GGA. The curves generated with the traditional and the new
methods are undistinguishable.}
\label{figpseu}
\end{figure}
\begin{figure}
\epsfig{figure=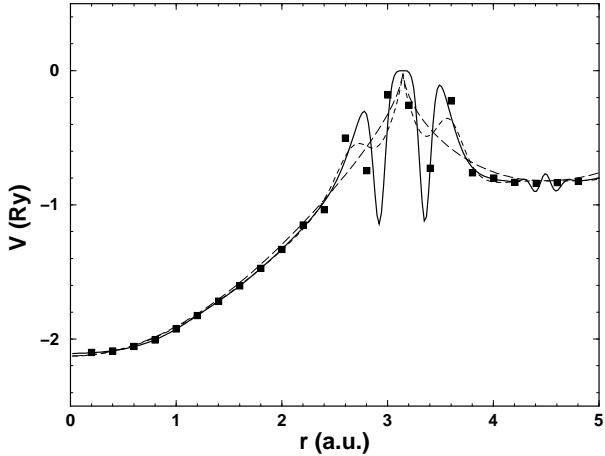,width=85mm}
\caption{
   XC potential for a spherical electron density with the form
$\rho({\bf r}) = (\sin r/r)^2$.
   The dashed line is the CA-LDA, the dotted line is the PBE-GGA and the
solid line is the PW91-GGA, all accurately calculated on a fine grid
with a step of 0.02 bohr. 
   The squares show that the PW91-GGA, calculated with the new method 
on a coarse grid (step = 0.2 bohr) have small deviations with respect 
to the fine grid.
   However, in fact these deviations tend to smooth out the large 
pathological wiggles developed by the PW91-GGA functional in regions 
of zero density or zero density gradient.
   This makes the present method very well behaved even with
very coarse grids.}
\label{figsinr}
\end{figure}

   To illustrate the method we have chosen three simple situations.
   Fig.~\ref{figpseu} shows the {\it s}-pseudopotential for Si,
with the Troullier-Martins recipe\cite{TM} and a core radius
of two bohr, calculated with the PBE-GGA and PW91-GGA functionals
using both our method (Eq.~(\ref{eqWB}))
and the usual method (Eq.~(\ref{eqvxcgga})), with the standard
radial grid used by the pseudopotential generation program.
   The curves for the two computational methods agree within $10^{-4}$ Ry 
and are indistinguishable in the scale of the figure. 
   The PW91-GGA pseudopotential has some wiggles that are due to
the term proportional to $\exp(-100 s^2)$ where
$s=\nabla \rho /(2 k_{\rm F} \rho)$, $k_{\rm F} = (3 \pi^2 \rho)^{1/3}$,
in the parameterization of the GGA exchange. 
   Those wiggles occur near the extrema of the valence electron density,
and they are not present in the PBE-GGA pseudopotential, which 
uses a different parameterization of the GGA exchange.
   By construction the screened pseudopotential is smooth,
without wiggles, so the appearance of those wiggles are
due to the unscreening of the pseudopotential.
   To see any difference between the usual and present methods,
we must use very coarse radial meshes.
   This is shown in Fig.~\ref{figsinr}, where we compare the XC potential 
for a model density $\rho({\bf r}) = (\sin r/r)^2$. 
   The PZ-LDA (dotted thin line), PW91-GGA (solid thick line), and 
PBE-GGA (dashed line), were first calculated on a fine radial mesh.
   Again there is no difference between the old and new methods
on the scale of fig.~\ref{figsinr} for the fine mesh. 
   However, recalculating
the PW91-GGA exchange  and correlation potential on a much coarser mesh
with the new method (squares)
one starts to see small differences in the potential values
in the regions where the PW91 parameterization has wiggles.
   The eleven point Lagrange formula for coarse sampling spans
to a region of 2 bohr, almost half of the horizontal
range of Fig.~\ref{figsinr}. 
   It misses the small wiggles around $r=3 \pi/2$ where $\nabla \rho = 0$ 
and $\rho \neq 0$, but it finds wiggles at $r=\pi$ where both 
$\nabla \rho = 0$ and $\rho = 0$.

\begin{figure}
\epsfig{figure=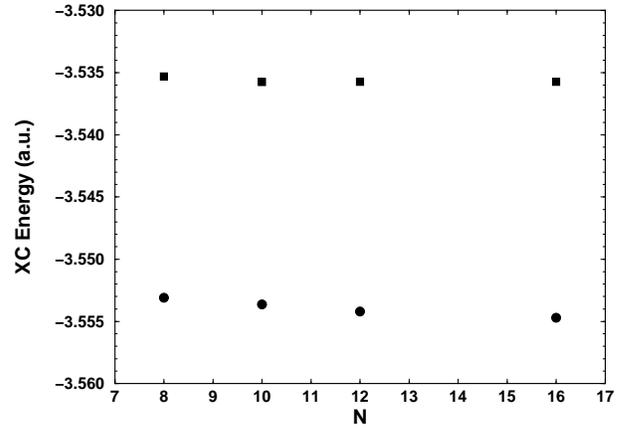,width=85mm}
\caption{
   XC energy in diamond as a function of the grid size $N \times N \times N$.
   Squares are from LDA and circles from GGA. 
   The charge density $\rho({\bf r})$ is the same in all cases.
   Only the sampling grid and the XC functional change.}
\label{fig_exc}
\end{figure}

   Our last example uses the pseudo-charge density of diamond. 
   A well converged plane wave expansion of the density is used to 
calculate $\rho_i$ for different $N \times N \times N$ grids. 
   Fig.~\ref{fig_exc} shows that the calculated XC energy is very 
stable with respect to grid size: even a $8 \times 8 \times 8$
grid gives a value accurate within $10^{-3}$ Hartree.
   The XC component of the stress also converges very fast,
and the method is very stable.
   In practice, the PW cutoffs required for a 
good convergence of the total energy impose larger 
grids than those needed to converge $E_{xc}$ with 
our method.
   Thus, the standard grids used in typical PW 
calculations are more than sufficient for $E_{xc}$.

\section{\bf Conclusions}

   We have derived formulas for the GGA-XC energy, potential, and stress,
as a function of the electron density given in a spatial mesh of points,
which may be unevenly distributed.
   Density gradients need not be provided, since they are calculated 
numerically using the density at the grid points, what leads to well
behaved formulas.
   The use of a unique definition of the gradients ensures an 
{\em exact consistency} between the calculated values for the energy,
potential, and stress, including all corrections for basis set
incompleteness and changing grid points.
   As the number of grid points increases, we recover the exact results
obtained using the virial theorem.

\acknowledgements

   This work was supported by the Fundaci\'on Ram\'on Areces and 
by MCT/DGI grants PB00-1312 and PB98-0368-C02-01.


\begin{references}
\bibitem{notation}
For convenience, we will place indexes and labels indistinctly as sub or 
superscripts, without meaning a tensorial notation nor implicit summations.
Indexes $i,j$ are used for the $M$ mesh points, $\alpha, \beta, \gamma$ 
for the three cartesian coordinates, $\mu, \nu$ for the three lattice or mesh
coordinates, $a,b,c$ for the $N$ basis functions, and $n,m$ for the $N$ 
hamiltonian eigenstates.
\bibitem{sign}
   We use here the convention that the stress has the opposite sign than
the pressure.
   Often the opposite convention is used, so the reader may
find an overall change of sign in the expressions
given by different authors.
\end{references}
\end{document}